\documentstyle[11pt,aaspp4]{article}
\slugcomment{To appear in Astronomical Journal}

\lefthead{Keller, Bessell, Da Costa}
\righthead{Young Magellanic Cloud Clusters II}

\begin{document}
 
   \title{Young Clusters in the Magellanic Clouds II}

\author{Stefan C. Keller$^{1}$, G. S. Da Costa \& M. S. Bessell}
\affil{Research School of Astronomy and Astrophysics, Institute of Advanced
Studies, The Australian National
University, Private Bag, Weston Creek P.O., ACT 2611, Australia.\\
Email: stefan, gdc, bessell@mso.anu.edu.au}
\altaffiltext{1}{Present address: Lawerence Livermore National Laboratory, L-413, 7000 East Ave. Livermore, 94550, U.S.A.}
\altaffiltext{2}{Based on observations with the NASA/ESA {\em Hubble Space
Telescope}, obtained at the Space Telescope Science Institute, which is
operated by the Association of Universities for Research in Astronomy, Inc.,
(AURA), under NASA Contract NAS 5-26555.}

\begin{abstract} 
We present the results of a quantitative study of the degree of extension to
the boundary of the classical convective core within intermediate mass
stars. The basis of our empirical study is the stellar population of four
young populous clusters in the Magellanic Clouds which has been detailed in
Keller, Bessell \& Da Costa (\cite{kel00}). The sample affords a meaningful
comparison with theoretical scenarios with varying degrees of convective core
overshoot and binary star fraction. Two critical properties of the population,
the main-sequence luminosity function and the number of evolved stars, form
the basis of our comparison between the observed data set and that simulated
from the stellar evolutionary models. On the basis of this comparison we
conclude that the case of no convective core overshoot is excluded at a 2
$\sigma$ level.
 
\end{abstract}
 
\keywords{Clusters:open(NGC 330, NGC 1818, NGC 2004, NGC 2100) - Magellanic
Clouds - Stars:evolution }
 
\section{Introduction}

The young populous clusters of the Magellanic Clouds (MCs) are amongst the
richest, resolvable young clusters in the Local Group. The clusters provide a
statistically significant sample of stars with which to constrain the
evolutionary behaviour of massive stars. This paper is the second in a series
detailing our study of four young populous clusters, NGC 330, 1818, 2004 \&
2100. These clusters have ages of 15-50 Myrs and main-sequence terminus masses
in the range of 9-12$\rm{M}_{\odot}$. In this mass range that the
treatment of convection, particularly within the convective core, is critical
to the course of stellar evolution. Ongoing debate focuses on the degree of extension of the convective core beyond that predicted by
standard, non-rotating stellar evolutionary models. This paper seeks to
provide quantitative constraints on the magnitude of convective core extension
through confrontation of our observations with the predictions of stellar
evolutionary models. Details of our observations are given in (Keller, Bessell
\& Da Costa 2000, hereafter Paper 1).

Extension of the convective core has traditionally been discussed in terms of
the formalism of convective core overshoot. Convective core overshoot (CCO) is
one easily visualised (and parameterised!) mechanism for extending the size of
the convective core. Under this description, gas packets rising in convective
motion from the core reach a distance from the centre where the acceleration
imparted to the convective element by the buoyancy force vanishes. This height
defines the classical Schwarzschild convective core boundary. Such packets
arrive at this height with non-zero velocity, hence some degree of overshoot
must occur. However, the classical picture considers that this degree of
overshoot is negligible (Saslaw \& Schwarzschild \cite{sas65}).

Numerous studies have attempted to ascertain the efficiency of CCO from a
theoretical basis with different results, ranging from negligible to
substantial (see e.g. Bressan et al.\ \cite{bre81}). An analytical approach appears
limited given the complexity of the phenomenon. Laboratory fluid dynamics
shows that an understanding of mixing requires a description of the turbulence
field at all scales, a problem which will require detailed hydrodynamical
modelling.

Recourse to observation of stellar populations is required to ascertain the
amount of convective core extension to apply within stellar evolutionary
models. Demonstration that the convective core is of greater extent than the
classical core does not, however, determine its causation. Several
effects can bring about an extension of the convective core above that
predicted by classical models. Opacity, for instance is critical in
determination of the convective core size. The superseding of the old LAOL
opacities (Huebner et al.\ \cite{hub77}) by those of OPAL (Iglesias et
al. \cite{igl92}) resulted in a substantial increase in the size of the
classical convective core (see Stothers \& Chin \cite{sto92}). Further, recent
studies indicate that rotation can provide a natural way to bring about
increased internal mixing and hence a larger core (Heger \& Langer
\cite{heg00} and Meynet \& Maeder \cite{mey00}). Nevertheless, the formalism
of CCO offers a straightforward parameterisation of the effect of convective
core extension regardless of causation. For this reason the following
discussion is in terms of CCO.

CCO brings about a number of important evolutionary changes which are
expressed in the colour-magnitude diagram (CMD) of a cluster population. In
particular, the MS lifetime is extended and stars evolve on the main-sequence
(MS) to cooler temperatures and higher luminosities relative to non-overshoot
models. Further, the subsequent post-MS evolution occurs at a more rapid pace,
leading to a number ratio of He burning stars to H burning stars that is
strongly dependent on the degree of overshoot ($\Lambda_{c}$; Langer \& El Eid \cite{lan86}, Bertelli et al.\ \cite{ber86}). 

Through comparison of the observed cluster CMD with the predictions of
stellar evolutionary models containing various degrees of CCO we are
able to ascertain the amount of overshoot exhibited by the
population.  In considering massive stars, however, it is necessary
to examine young clusters, which in the case of our Galaxy are too
poorly populated to individually yield much insight. It is possible
to combine data from sets of similar aged clusters as in the work of
Mermilliod \& Maeder (\cite{mer86}) and Maeder \& Mermilliod
(\cite{mae81}), yet such sets are inevitably heterogeneous in age and
metallicity. The young populous clusters of the Magellanic Clouds
contain upward of 10$\times$ the mass of their Galactic counterparts
and thus provide ideal targets for study.

Perhaps the best studied system is NGC 1866 (age=100-200Myr; Testa et
al. \cite{tes99}). The intermediate age population of this system has been
considered the proof both for (Becker \& Mathews \cite{bec83}, Chiosi et al.\
\cite{chi89}) and against (Brocato et al.\ \cite{bro94}, Testa et al.\
\cite{tes99}) CCO. In Chiosi et al.\ (ibid) the degree of CCO exhibited by the
population on NGC 1866 is estimated at $\Lambda_{c}$=0.5. Whilst remaining a
contested issue, this degree of overshoot then forms the basis of the physical
input supplied to the standard evolutionary models (SEMs) of the Padua group
(e.g. Fagotto et al.\ \cite{fag94}). The motivation for the present study is
to extend the number of young clusters which have been examined in regard to
the amount of CCO and to place the need or otherwise for CCO on a firmer
basis.

Previous ground-based and IUE studies of the clusters of the present work have
revealed several features which are anomalous compared to SEMs which do not
include overshooting. These features are indicative of some degree of
convective overshoot within the population. However, further conclusions from
these studies are limited since the observations have been restricted to the
brightest members, the only stars for which accurate effective temperatures
were attainable. They extend to just below the MS turnoff. In addition
previous studies have made use of reddening, metallicities, and in some cases
opacities, which have since been superseded by more appropriate input. The
present paper revisits and forms a significant extension to these previous
studies. We draw together new data presented in Paper 1 together with a set of
stellar evolutionary models containing homogeneous input physics.

\section{Observations and Data Reduction}

\begin{figure}
\plotone{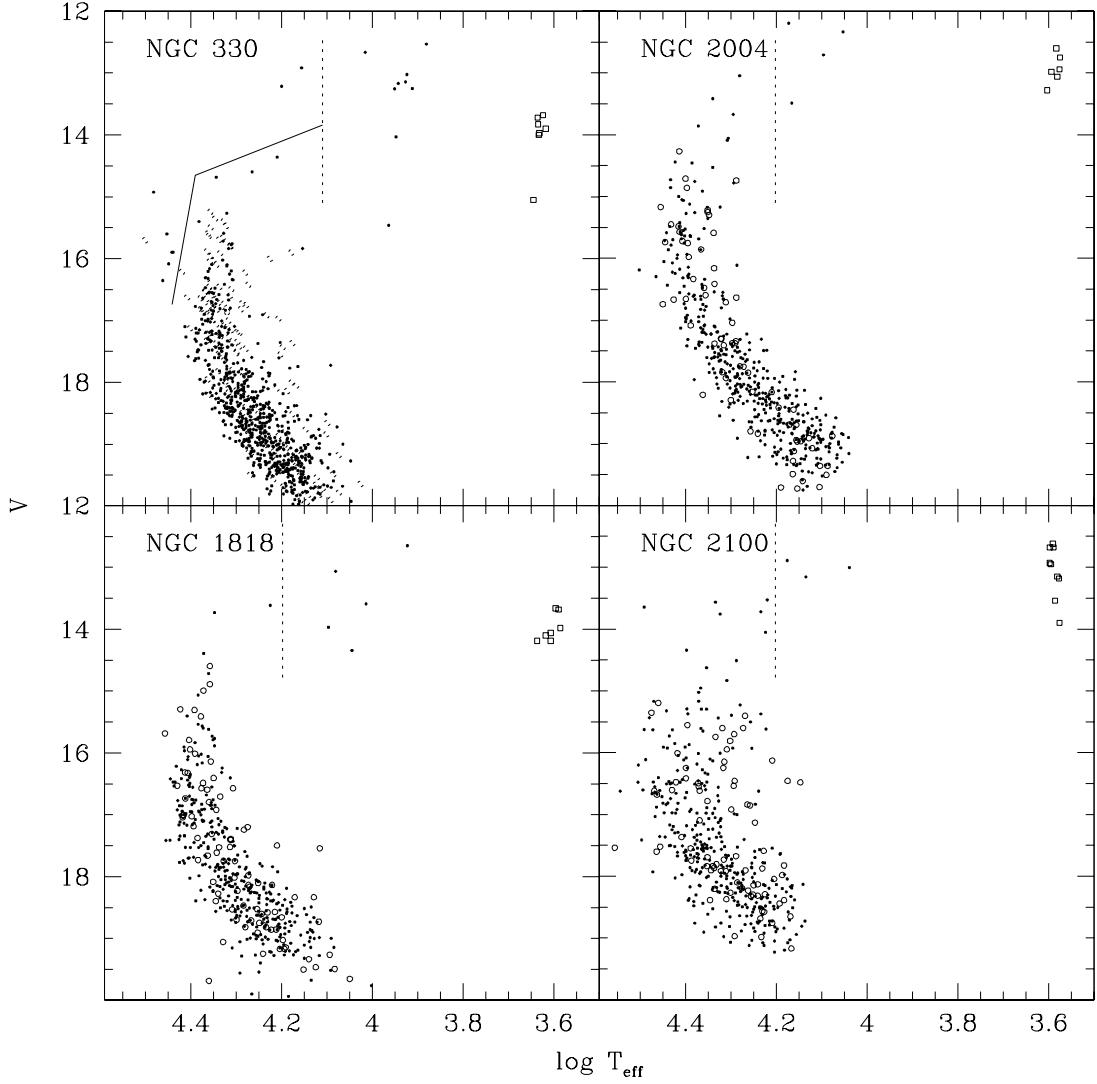}
\caption{The V,$T_{eff}$ diagrams of the four clusters of the sample; top left
NGC 330 (those objects above the solid line are excluded - see text) , bottom
left NGC 1818, top right NGC 2004 and bottom right NGC 2100. The post-MS
population are those objects to the right of the dashed line. Be stars are
shown as open circles. Red supergiants from Keller (\cite{kel99ir}) are shown
as open boxes.}
\label{hrds}
\end{figure}

The data presented in this paper were obtained with the {\it{Hubble Space
Telescope}} Wide Field and Planetary Camera 2 (WFPC2).  Exposures in F555W
($\equiv V$), F160BW ($\Delta lambda=446$\AA, $\lambda_{eff}=1491$\AA) and
F656N filters were obtained. More details on the reduction of this data can be
found in Paper 1. We delimited two important regions for
the analysis to follow: the MS region and the evolved supergiant region. These
regions are shown on the cluster HR diagrams shown in Figure \ref{hrds}.

\placefigure{hrds}

In the case of NGC 330 we have excluded those stars which have been described
as potential blue stragglers (see Paper 1) and stars A01 and B37 (using the
nomenclature of Robertson~\cite{robo}) which are significantly brighter than
the MS terminus (by two magnitudes in $V$). These stars may represent members
of the blue straggler population at a later stage of MS evolution.

A particularly useful comparison between model predictions and observations is
achieved through construction of the MS luminosity function. For this reason
particular attention is given to the treatment of photometric completeness and
field contamination in the observed sample.

\subsection{Photometric Completeness} 
\begin{figure}
\plotone{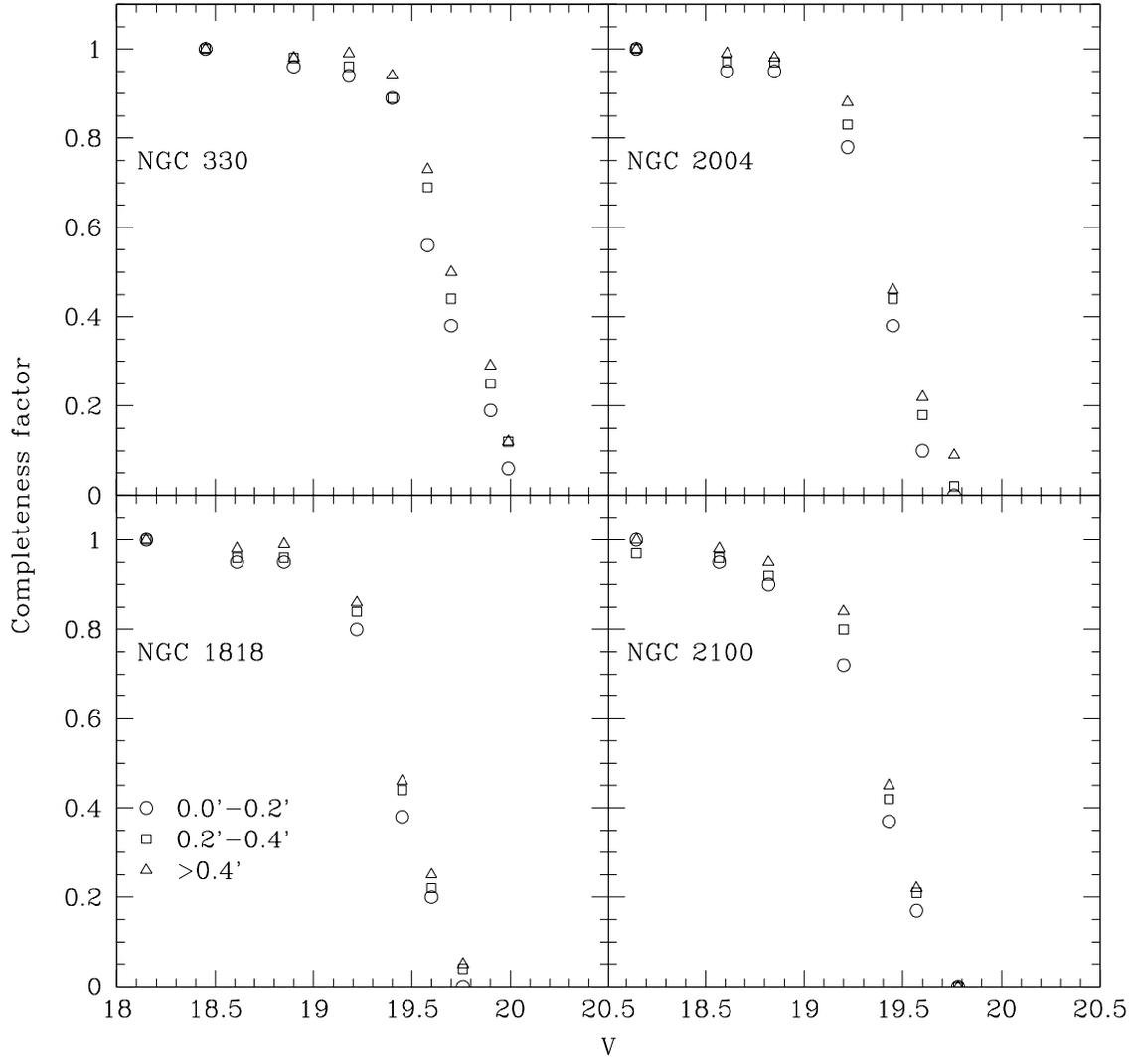}
\caption{Completeness in F555W ($\equiv V$) for the sample on the MS within the four clusters as a function of radial distance from the cluster centre and apparent magnitude.}
\label{completeness}
\end{figure}

Quantitative estimation of the degree of completeness was established through
extensive artificial star tests. A set of artificial stars was generated
within each of the 4 frames of the WFPC2 images with a range of magnitudes. An
identical reduction procedure was performed on the enriched frames to that
performed on the original frames. An artificial star was considered as
``recovered'' if the recovered image centroid agrees with the actual position
to within 1px and if the recovered magnitude is within 0.2 mag of the actual
magnitude. The completeness factor is the ratio between the number of
artificial stars recovered to the number of stars originally simulated. The
completeness of the present photometry is determined by the F160BW exposures
as these have significantly lower S/N than the F555W. Using the above
procedure, we have measured the completeness within our F160BW frames. We have
then used a spline fit to the locus of the MS in the F555W, F160BW$-$F555W CMD
to transform the completeness in F160BW to the corresponding completeness in
F555W {\it{on}} the MS. This is shown in Figure \ref{completeness}. The
completeness is a moderate function of radius but mostly a function of
magnitude. For this reason, our completeness correction is applied in three
radial regions.

\placefigure{completeness}

\subsection{Field Star Contamination}

Our imaging has not included a separate field sample for the purposes of field
subtraction. For determination of the field star contamination we have relied
on previous $V$, $\;I$ photometry detailed in Keller, Wood \& Bessell
(1998). In this previous study, a field 10\arcmin$\;$ square centred
on the cluster was imaged. An examination of the number density of stars
clearly delimits the apparent extent of the cluster. The field star
contribution within each field was then established from the number of MS
stars outside the apparent cluster radius. The number of post-MS stars
expected from the field sample with luminosities comparable with those arising
from the cluster population within the WFPC2 field (total area of
5.8\arcmin$^{2}$) is low, typically 0.1 star/field. The field contribution in
0.5 mag bins is subtracted at random from the completeness corrected sample.

In order to minimise the influence of the corrections for completeness and
field stars on the observed luminosity function, we define a limiting
magnitude to our sample at $V_{min}$ (see Table~\ref{x}). At this magnitude
the completeness is always more than 80\% and the field contribution no more
than 10\% of the uncorrected sample. Table~\ref{x} presents a summary of our
sample.

\placetable{x}

\section{Previous Studies of the Clusters}

Previous studies of the four target clusters have largely focussed on NGC 330
and NGC 2004. Such studies have been restricted to the upper portion of the MS
and present a comparison between the observed position of upper-MS and post-MS
stars and evolutionary models. The majority of previous studies have made use
of reddenings, metallicities, and in some cases distance moduli and opacities
which have since been superseded. The following is a brief review of how the
conclusions of previous studies stand in the light of current stellar
evolutionary models (SEMs) and observational data.

\subsection{NGC 330} 

The study of NGC 330 by Stothers \& Chin (\cite{sto92}) presented a discussion
based on the population of the blue and red supergiants. The historical
motivation behind studies such as Stothers \& Chin was to restore in the model
predictions the observed blue-loop behaviour of post-MS stars. This blue-loop
behaviour was severely curtailed in model predictions through the
introduction of the modern OPAL opacities.

In the study of Stothers \& Chin different mixing scenarios were
considered: semi-convection with Schwarzschild and Ledoux criteria and
convective overshoot. The conclusions of this study can be summarised as
follows: 1) the observed red supergiants within the cluster exclude the
Schwarzschild criteria, only with the Ledoux criteria can red supergiants be
formed at Z=0.004, 2) only models with little or no overshoot match the
effective temperature of the blue supergiants and the difference in average
luminosity between the red and blue supergiant stars. In a later paper
Stothers \& Chin (\cite{sto94}) go further to claim that the only way they can
get the blue-loop evolution of He burning supergiants to agree with
observations is if the OPAL opacities were to be increased; they called for an
increase in the opacity at the base of the outer convective zone of at least
70\%.

Langer \& Maeder (\cite{lan95}) discussed the inability of stellar
evolutionary models to account for the dependence on metallicity of
the number ratio of blue to red supergiants. This ratio is highly
sensitive to input physics and represents an unresolved problem. From
the data presented in Paper 1, the difference in average luminosity
between the blue and red supergiants is $\Delta$log L/$\rm{L}_{\odot}
= -0.10\pm0.10$. This is comparable to the estimate by Stothers \&
Chin who find $\Delta$log L/$\rm{L}_{\odot} = -0.24\pm0.08$. Models by
Fagotto et al.\ (\cite{fag94}; Z=0.004, $\Lambda_{c}$=0.5) for an
appropriate age of log(age)=7.5 predict $\Delta$log L/$\rm{L}_{\odot}
= -0.08$, similar models to with no overshoot which predict
$\Delta$log L/$\rm{L}_{\odot}=-0.12$. Evidently no distinction can be
made between these models on the basis of the average luminosity
difference between blue and red supergiants.

Caloi et al.\ (\cite{cal93}) arrived at a provocative conclusion from their
study of IUE spectra of the brightest blue stars in NGC 330. They found that
stars of the upper MS {\it{avoid}} the regions of the HR diagram predicted to
be populated by the evolutionary models of Brocato \& Castellani
(\cite{bro93}) (Z=0.003, LAOL opacities \& no core overshoot). The observed
stars lie cooler that the theoretical MS and hotter than the predicted
position of the He burning supergiants.

Chiosi et al.\ (\cite{chi95}) demonstrated that models with OPAL opacities and
overshoot (both core and envelope) significantly improved the match between
model and observation. They stopped short of drawing conclusions about
internal convection from the cluster population and were unable to fit the
cluster CMD (based upon $B$$-$$V$ colours) with a single age, requiring
instead a large age spread (log age=7.0-7.7) between the MS and supergiant
stars. In Paper 1 we demonstrated that such an age spread is inconsistent with
our observations. Rather we found that the age of NGC 330 is constrained on
the basis of isochrone fits to log age=7.5$\pm$0.1 (on basis of isochrones by
Fagotto et al.\ \cite{fag94}). The surmised large age spread of Chiosi et
al. is the result of a combination of the inherent insensitivity of the
$B$$-$$V$ colour for stars with temperatures such as those associated and with
the upper MS of these clusters and the addition of a Be star component,
members of which have discrepant $B$$-$$V$ colours (Keller, Wood \& Bessell
\cite{kel99}).

\subsection{NGC 2004}
Caloi \& Cassatella (\cite{cal95}) presented an IUE-based study of the upper
MS of NGC 2004. They arrived at a similar conclusion to that of their study of
NGC 330 discussed above. Again they found that the stars of the upper MS form
a vertical sequence at cooler temperatures than that predicted by
evolutionary models (Z=0.003, LAOL opacities \& no-overshoot). In Paper 1 we
showed that a model with a choice of metallicity in closer agreement with
determinations and with OPAL opacity , together with a level of core overshoot of 0.5, is in good agreement with the temperatures of the stars on the upper MS.

Caloi \& Cassatella pointed out that the luminosities of the red
supergiant population are below the luminosity of the MS terminus. In their
study they relied upon temperatures determined from $B$$-$$V$ colours. From
our IR photometry (Keller \cite{kel99ir}) we found that the temperatures of
this sample have been systematically underestimated by 400K. The resulting
change in the bolometric correction brings the luminosity of the red
supergiants to the level of the MS terminus.

Subramaniam \& Sagar (\cite{RAM95}) examined five young MC clusters
including NGC 2004. They examined the normalised integrated luminosity
function (NILF), which is defined at a point $V_{bin}$ as the sum of
the number of stars brighter than $V_{bin}$ divided by the total
number of post-MS stars: \(\sum_{}^{V=V_{bin}} N / N_{PMS}\), where
$N_{PMS}$ is the total number of post-MS stars.They compared the NILF
produced by a series of SEMs, with and without overshoot and with LAOL
and OPAL opacities. They were unable to decide on a favoured model
because they treated the present-day mass function slope as a free
parameter. They do remark that the models incorporating overshoot do
manage to best reproduce the observed features of the CMD.

\section{A Comparison Between Stellar Evolutionary Models and Observation}

\subsection{The Stellar Evolutionary Models}

\begin{figure}
\plotone{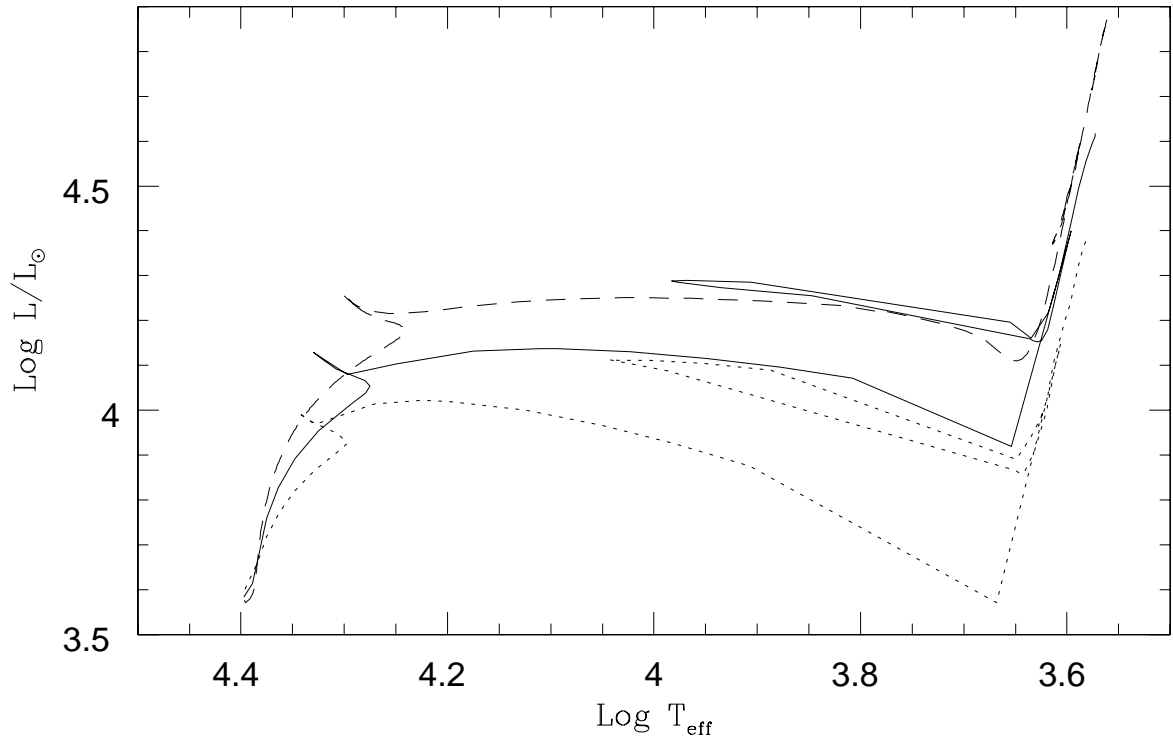}
\caption{Evolutionary tracks for a 9$\rm{M}_{\odot}$ star of Z=0.008 and $\Lambda_{c}$ = 0.00 (dotted line), $\Lambda_{c}$ = 0.50 (solid line) and $\Lambda_{c}$ = 1.00 (dashed line).}
\label{trackfig}
\end{figure}

Table \ref{models} lists the stellar evolutionary models used in the
present work. Models 1,2,4 and 5 are taken from (Fagotto et al.\
\cite{fag94} \& Bertelli \cite{ber94}). Models 3 and 6 featuring
extreme overshoot were kindly computed by the Padova group. In this
way we have established two uniform sets of models of appropriate
metallicity, whose input physics differs only in the degree of
CCO. Models of the Geneva group use a different parameter definition
for expressing the same amount of CCO ($\Lambda_{c}$ =
2.$d_{ov}/\rm{H_{P}}$ where $d_{ov}/\rm{H_{P}}$ is the amount of CCO in
the Geneva group models). Figure \ref{trackfig} shows evolutionary
tracks from models \#4, 5 \& 6 for a 9$\rm{M}_{\odot}$ star.

\placetable{models}

The blue loops in models \#3 \& 6 are unrealistically short. This detail does
not pose a problem for the present study because in the analysis we use only
the total number of evolved stars, independent of their position in the HR
diagram.

Synthetic cluster populations were derived from the above evolutionary
tracks by means of our own code which incorporates photometric errors,
the effect of crowding, the presence of binary systems with an
arbitrary mass ratio, an age spread within the population and an
arbitrary present-day mass function (PDMF). Synthetic CMDs can be
produced over an age range from a few Myrs to several Gyrs.

\placefigure{trackfig}

Here we focus on two joint constraints: (1) the number of stars per unit of
luminosity on the MS (the luminosity function) and (2) the number of post-MS stars. Model predictions of both must be concordant with the observations to merit an acceptable match. 

\subsection{ The Main-Sequence Integrated Luminosity Function}

Studies of the luminosity function typically make use of the normalised
integrated luminosity function. Such an approach is valid in
these previous studies where the data set does not include a significant
sample of the MS. In our case we have a well-defined (and essentially
complete) MS down to at least 3 mag. below the MS terminus. We can adopt the
number of stars within a portion of the MS as the normalisation factor instead
of the number of post-MS stars. In this way, our synthetic CMDs are populated
until the observed total number of MS stars brighter than
$\rm{M}_{V}<$$V_{min}$ is reached. This represents a superior normalisation
factor since the number of MS stars in this range is much larger than the
number of post-MS stars and is therefore less affected by stochastic
fluctuations. The total number of stars within this range does however depend
on the adopted distance modulus, the slope of the mass function and the
fraction of binary stars present within the cluster. We now discuss each of
these in turn.

\subsubsection{Distance moduli}

The distance moduli of the Magellanic Clouds are the topic of much debate. We have taken a distance modulus of 18.45 to the LMC and 18.85 to the SMC. In the review of van den Bergh (1998) the likely uncertainty in such estimates of the distance modulus are $\pm0.1$ mag.

\subsubsection{Present-day mass function}

We choose a slope of the PDMF equivalent to the Salpeter MF,
$\alpha$=-2.35. Originally derived from the solar vicinity, the
Salpeter MF has been shown by many recent studies to be appropriate to
a large range of environments and metallicities. Kroupa (\cite{kro00})
presents a review of determinations of $\alpha$, in particular over a
range of masses. We shall not consider variation in the present-day MF
from the Salpeter value in the discussion to follow. However, let us
remark here that a decrease in $\alpha$ requires an older age to match
the data. For example, in the case of NGC 330 (Model \#2), changing
$\alpha$ to -2.0 requires an increase in age of the best-fitting model
of log age=0.05 dex.

\subsubsection{Binary fraction}

Elson et al.\ (\cite{els98}) find that NGC 1818 has a conspicuous population
of binary systems. From direct observation of the binary star sequence they
estimate the binary fraction at 35$\pm$5\% within the cluster core radius and
25$\pm$5\% outside this radius. It is not possible from our data to
independantly determine the binary fraction. For this reason, the present
study considers two cases; no binaries and a spatially uniform binary fraction
of 30\% (section~\ref{binsection}) within each cluster.

Within our synthetic CMD code, the presence of binaries is accounted for as
follows: the mass of the star is generated at random from a probability
density function determined by the present-day mass function, the mass of the
secondary is then determined assuming a gaussian distribution of masses
centred on the mass of the primary. In this case we chose a narrow gaussian
distribution ($\sigma$=0.1M$_{\odot}$), in order to maximise the effects of
the binary population. Clearly, the effect on the luminosity of the system is
maximised if the two stars have similar masses.

\section{Discussion}

\begin{figure}
\plotone{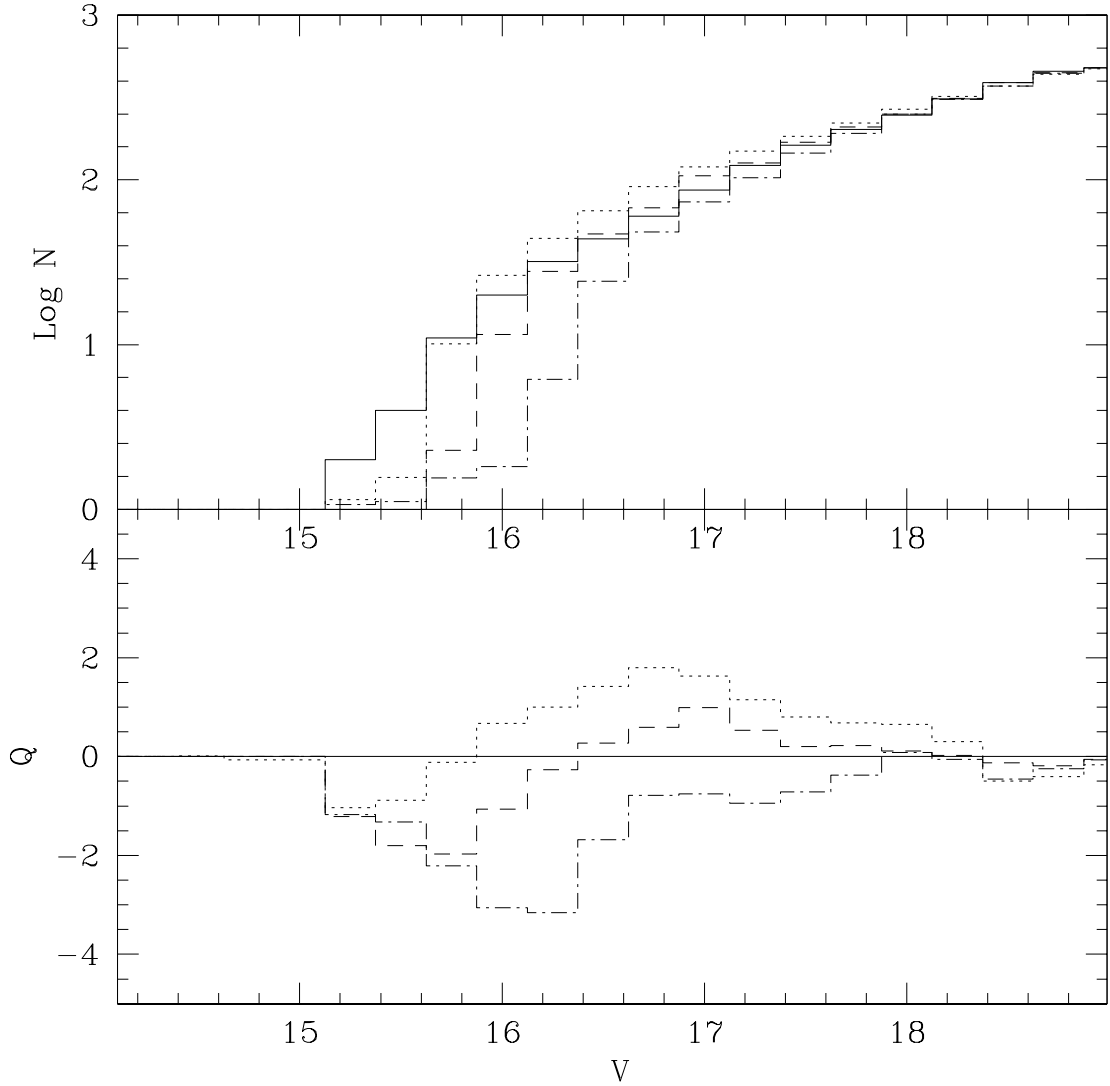}
\caption{Top. Comparison between the observed MS integrated luminosity function (MSILF) of NGC 330 (solid line) and model predictions for ages 7.30 (dotted), 7.36 (dashed line) and 7.50 (dot-dashed line) from Model \#1. Bottom. Plot of $Q$ for the above three ages. The model of log age=7.36 produces the minimum sum of residuals and hence represents the best fit to the MSILF for Model \#1.}
\label{figsig0.00}
\end{figure}

In the analysis to follow we will focus of the MS integrated
luminosity function (MSILF) for the purpose of comparing model with
data. The implications of the effective temperature information is
examined separately in Section \ref{efftemp}. The MSILF is formed for
each magnitude ($V_{bin}$) by the summation of the number of objects
with $V<V_{bin}$ (top panel of Figure~\ref{figsig0.00}). The
comparison between Model \#1, incorporating no extension of the
convective core, and NGC 330 is summarised in Figure~\ref{figsig0.00}
for three ages: at log age=7.30, 7.36 and 7.50. In order to examine
the difference between the MSILF observed and that simulated, we have
constructed the quantity $Q=(\rm{N_{obs}-N_{theo}}) /
(\sqrt{\rm{N_{theo}} + \rm{N_{obs}}})$ (seen in the bottom panel of
Figure~\ref{figsig0.00}). 

$Q$ is constructed in this manner in order to provide a clear
indication of where significant deviations occur in the model
predictions. The nature of the problem means that these deviations
occur towards the top of the MS, for this reason other well known
statistical quantifiers such as the K-S statistic are not well suited
to its description.  As the total number of stars brighter than
$\rm{M}_{V}<$$V_{min}$ is equal to the observed number, $V_{min}$ by
definition corresponds to $Q$=0. Moving towards brighter magnitudes,
the value of $Q$ will not necessarily remain around zero: the larger
the range over which $Q$ remains close to zero, the better the fit.

Consider the case of log age=7.30; here the MS terminus point is in close
agreement with that observed: however, the model does not mimic the observed
{\it{density}} on the MSILF. The model produces too high a density towards the
end of the MS, hence the positive values of $Q$. In the case of a log age=7.5,
the model fails to reproduce the MS terminus, which results in a region of
negative $Q$. The area, $A=\sum |Q|$, in the above figures gives a global
description of the goodness of fit for each model to the data. In the case of
Model \#1 the maximum likelihood is obtained for a log age of 7.36.

\placefigure{figsig0.00}

\begin{figure*}
\plotone{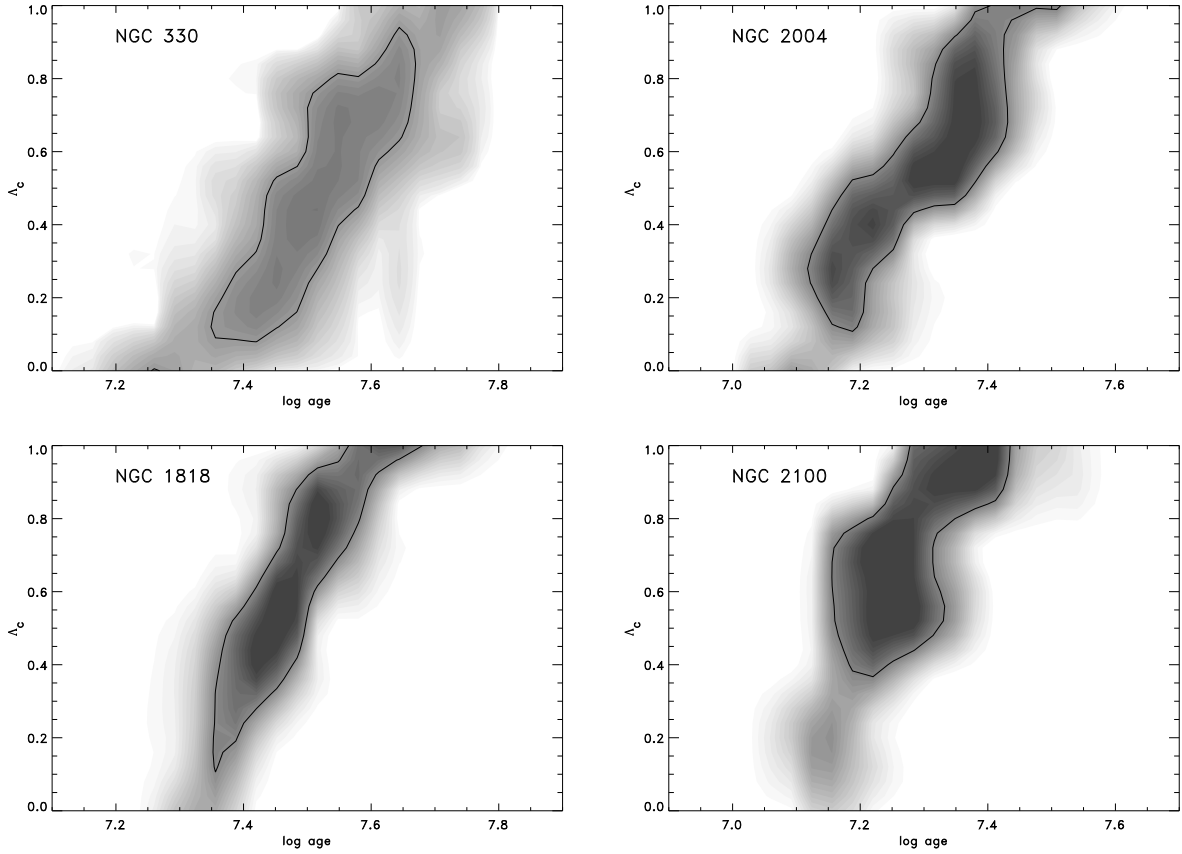}
\caption{Greyscale plot of probability density from the comparison of the observed MSILF with the model prediction for: NGC 330 (top left), NGC 1818 (bottom left), NGC 2004 (top right) and NGC 2100 (bottom right). A 1$\sigma$ contour level is shown.}
\label{lfs}
\end{figure*}

We have performed a quadratic interpolation between the three models in order
to sample the range of CCO from 0.0 to 1.0. This is justified as the effects
brought about by modification of the CCO level are well behaved. We have
repeated the determination of $A$ 100 times for each value of age to enable us
to describe the stochastic uncertainty involved. We repeat for values of CCO
in the range covered by our models. In the parameter space of each cluster
there exists a minimum, $A_{min}$, which represents the favoured model. We
express every other determination of $A$ in terms of the number of standard
deviations from $A_{min}$. The normalised probability density at each point is
then given by $L$=$A_{min}e^{-\sigma^{2}/2}$. The resulting probability
density is shown in Figures~\ref{lfs} as a greyscale plot with a 1$\sigma$
contour level overlaid.

We find that this method provides a more robust determination of the age of
the cluster than is available by a simple isochrone fit to the data. This
method makes use of the bulk of essentially unevolved MS members whose ILF
depends significantly on the assumed age. On the other hand the degree of
extension to the convective core is only weakly constrained, but in the case
of each cluster favours a non-zero level of CCO.

\begin{figure}
\plotone{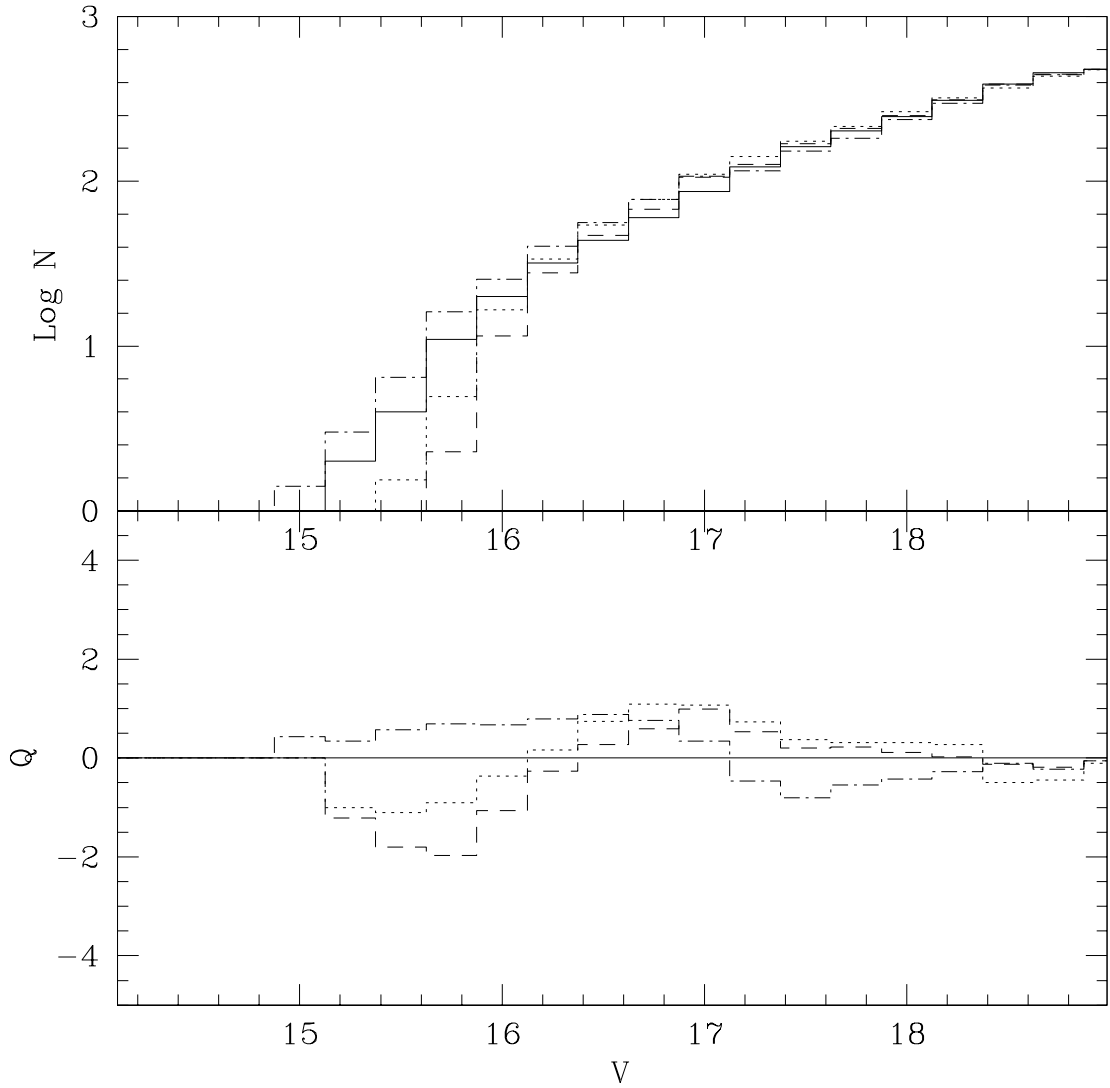}
\caption{Comparison of the best-fitting models to the MSILF of NGC 330 for
$\Lambda_{c}$=0.0 (dashed line), 0.5(dotted line) and 1.0(dot-dashed line). The
best-fit ages are (log age) 7.36, 7.52 and 7.74, respectively.}
\label{compbestfits}
\end{figure}

It is important to consider why the MSILF of these clusters is best fit by a significant level of CCO. A comparison of the best-fitting models to the MSILF of NGC 330 is shown in Figure~\ref{compbestfits}. The differences between the models of various levels of overshoot are due to a change in the density along the MS. Models with a larger degree of overshoot evolve towards higher luminosities in the final stages of MS evolution which results in a stretching out of the MSILF towards the MS terminus. The lower panel of Figure~\ref{compbestfits} shows this clearly. The $\Lambda_{c}$=0.0 model terminates too rapidly to mimic the observations, the $\Lambda_{c}$=1.0 model on the other hand terminates too slowly.

\placefigure{lfs}

\placefigure{compbestfits}

\section{The Post-MS Population}

The number of evolved stars predicted by the evolutionary models is another
important point of comparison with the observed cluster population. Since the
duration of post-MS evolution for a star of a given mass is determined to
first order by the He core size at the end of central H burning, the number of
evolved stars (${N_{evol}}$) which result from a given population size is very
sensitive to the level of CCO. Those with larger amounts of CCO will attain
larger He cores and will progress more rapidly through post-MS evolution. As a
consequence a smaller number of evolved stars will result from such a
population compared with a population incorporating less CCO.

\begin{figure*}
\plotone{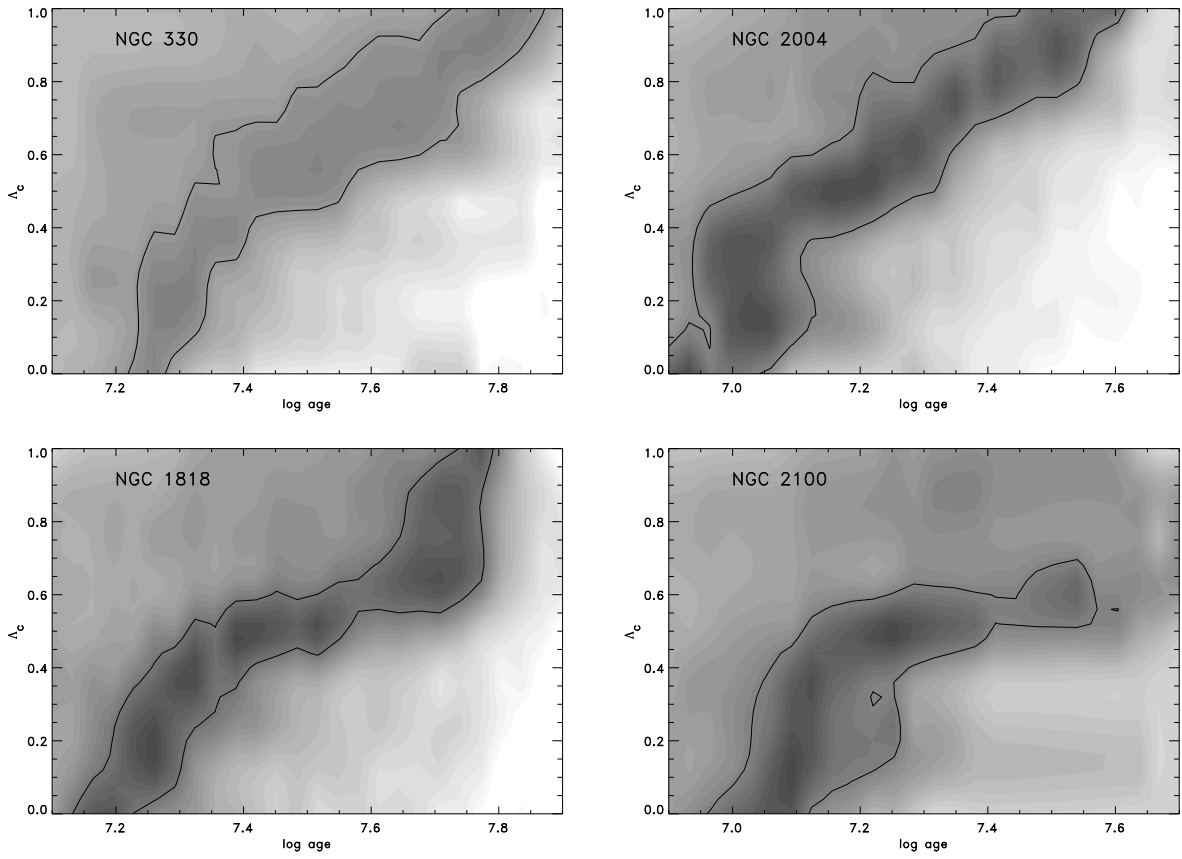}
\caption{Greyscale plot of $Q_{evol}$ for the four clusters. A 1$\sigma$ contour is shown.}
\label{evolfig}
\end{figure*}

\placefigure{evolfig}

We determine $Q_{\rm
evol}$=$(|{N_{evol}}^{o}-{N_{evol}}^{m}|)/\sqrt({N_{evol}}^{o})$, where
${N_{evol}}^{o}$ is the observed number of evolved stars (see Table~\ref{x})
and ${N_{evol}}^{m}$ that simulated. In this way, $Q_{\rm evol}$ expresses
how well the model predicts the number of evolved stars observed. The result
for each cluster is shown in Figure~\ref{evolfig}. Unlike the comparison of
the MSILF to the model output, this comparison does not constrain the level of
CCO rather it describes a locus in the (age, CCO) plane.

In each case the value of $Q_{\rm evol}$ asymptotes to
${N_{evol}}^{o}/\sqrt({N_{evol}}^{o})$ towards younger ages because
${N_{evol}}^{m}$ drops towards zero. On the other hand, towards older ages
${N_{evol}}^{m}$ becomes large. Within this range of ${N_{evol}}^{m}$ for each
$\Lambda_{c}$ there exists an age at which ${N_{evol}}^{m}$ corresponds to
${N_{evol}}^{o}$.

\subsection{Joint Constraints on the Degree of Internal Mixing}

\begin{figure*}
\plotone{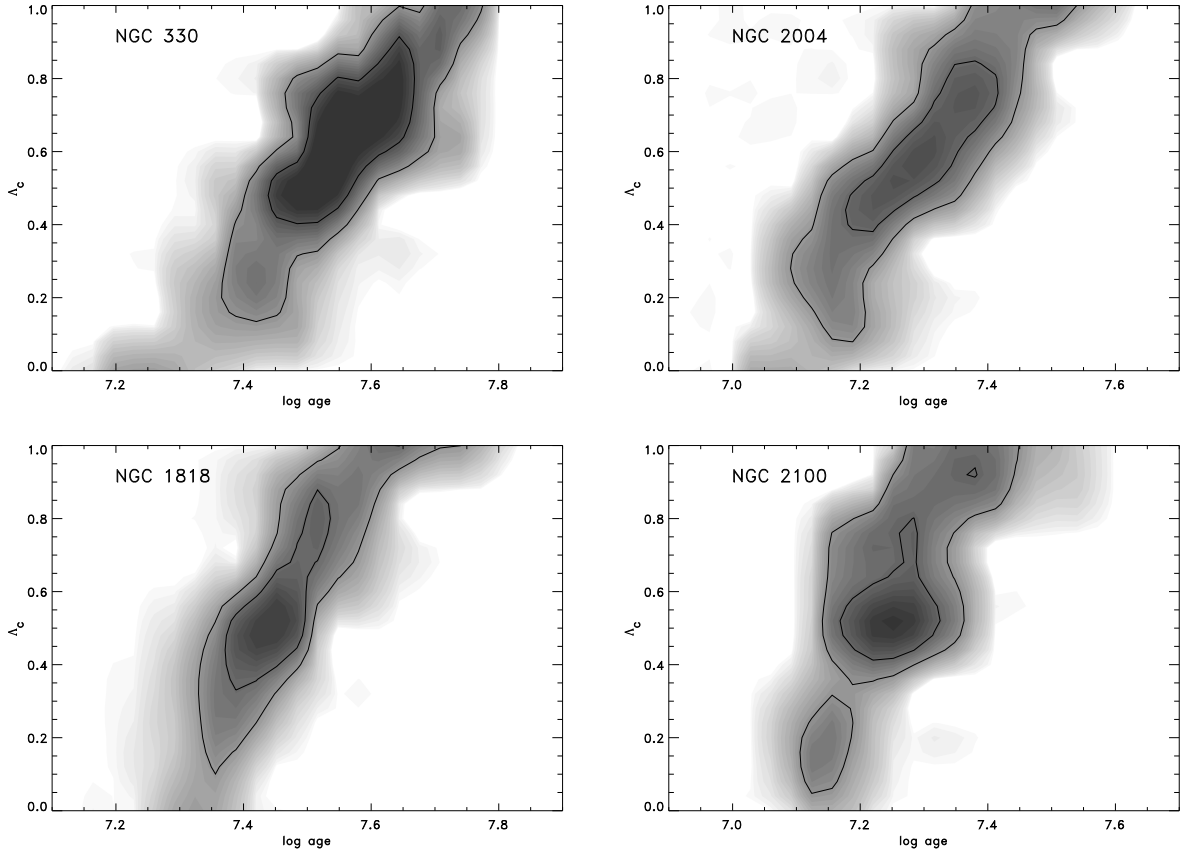}
\caption{Greyscale plot of the joint-probability density from the combination of
the MSILF and $N_{evol}$ constraints for each cluster. 1 and 2 $\sigma$ contour levels are shown.}
\label{total}
\end{figure*}

Figure~\ref{total} presents the joint constraints formed by the multiplication
of the probability density functions derived from $Q$ and $Q_{evol}$
discussed above. Ideally both constraints would be complementary i.e. they
would be in orthogonal directions. However, although this is not the case, the
combination of the two constraints significantly tightens the constraints on
age and level of CCO.

\placefigure{total}

For each cluster the best-fitting age and $\Lambda_{c}$ is given in Table
\ref{tab:agenobin}. We see no significant metallicity or mass dependence in
the best-fitting $\Lambda_{c}$ over the range of metallicity ([Fe/H]=-0.82 for
NGC 330: Hill 1999a to -0.50 for NGC 2004: Hill 1999b) and MS terminus mass
(M=9$\rm {M}_{\odot}$ for NGC 330 to 14$\rm {M}_{\odot}$ for NGC 2100). By simply combining the results from the four
clusters shows that a level of CCO $\Lambda_{c}$=(0.56$\pm$0.19) is
optimal. In each cluster the case of no-overshoot is ruled out to at least a 2
$\sigma$ level.

\placetable{tab:agenobin}

\section{Inclusion of a binary fraction}
\label{binsection}

\begin{figure}
\plotone{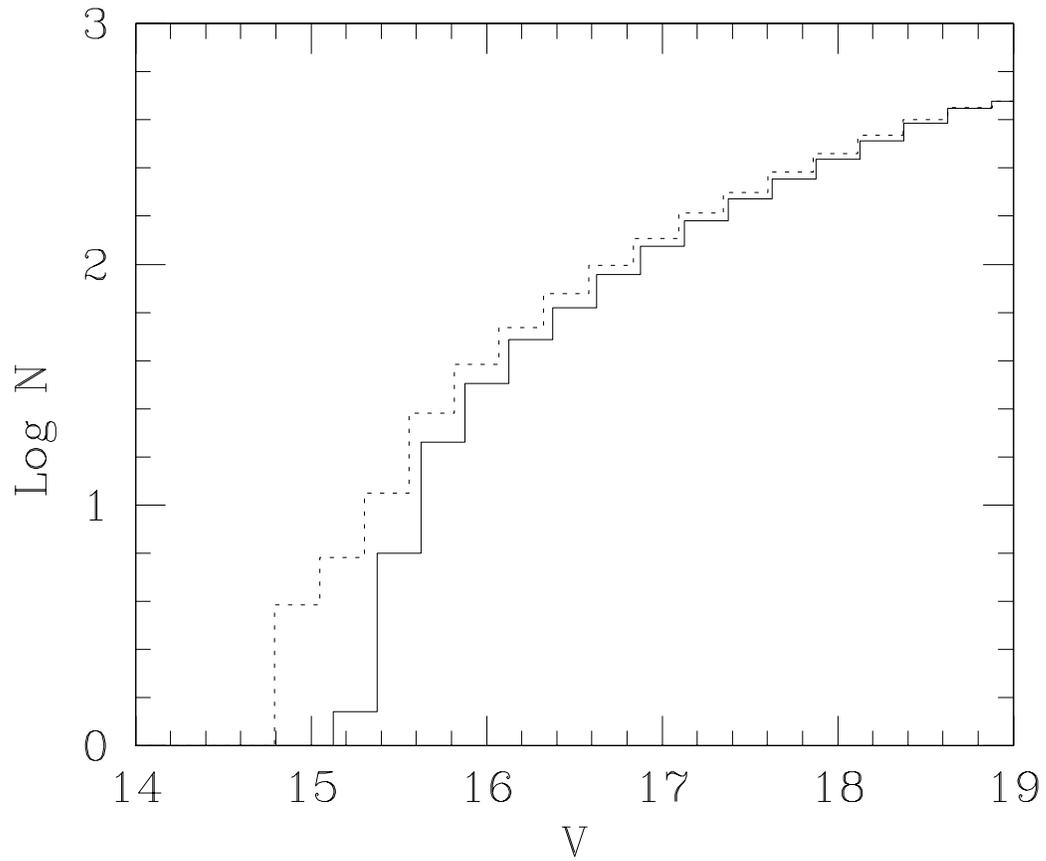}
\caption{Model (\#2:log age=7.50) MSILF with a 30\% binary fraction (dotted line) and without binaries (solid line). }
\label{binnobin}
\end{figure}

\placefigure{binnobin}

\begin{figure*}
\plotone{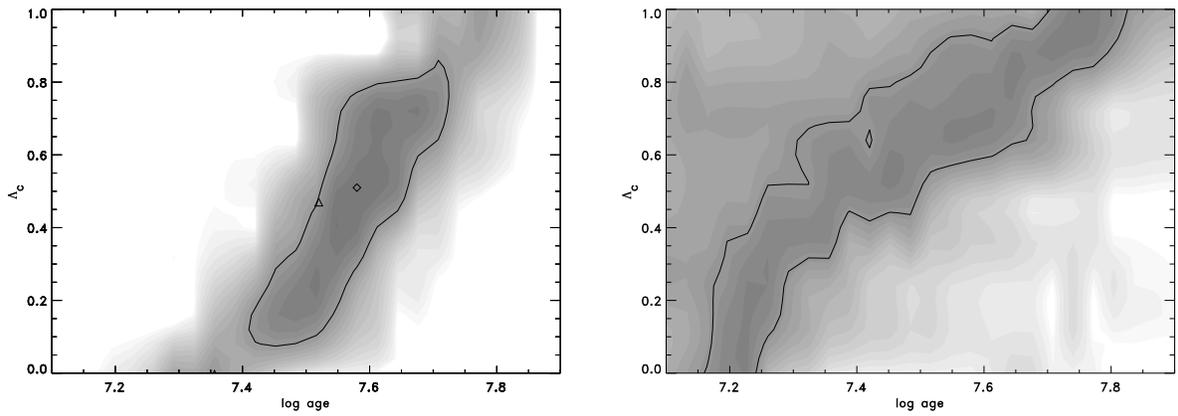}
\caption{{\bf{left}} Greyscale plot of probability density from the MSILF  (the
triangle symbol represents the best-fitting point from Figure
\ref{compbestfits}) for NGC 330
with a binary fraction of 30\%. {\bf{right}} The equivalent plot for the
evolved star probability density. A 1$\sigma$ contour is shown. Compare with
Figures \ref{compbestfits}\&\ref{evolfig}.}
\label{330bin}
\end{figure*}

\placefigure{330bin}

\begin{figure*}
\plotone{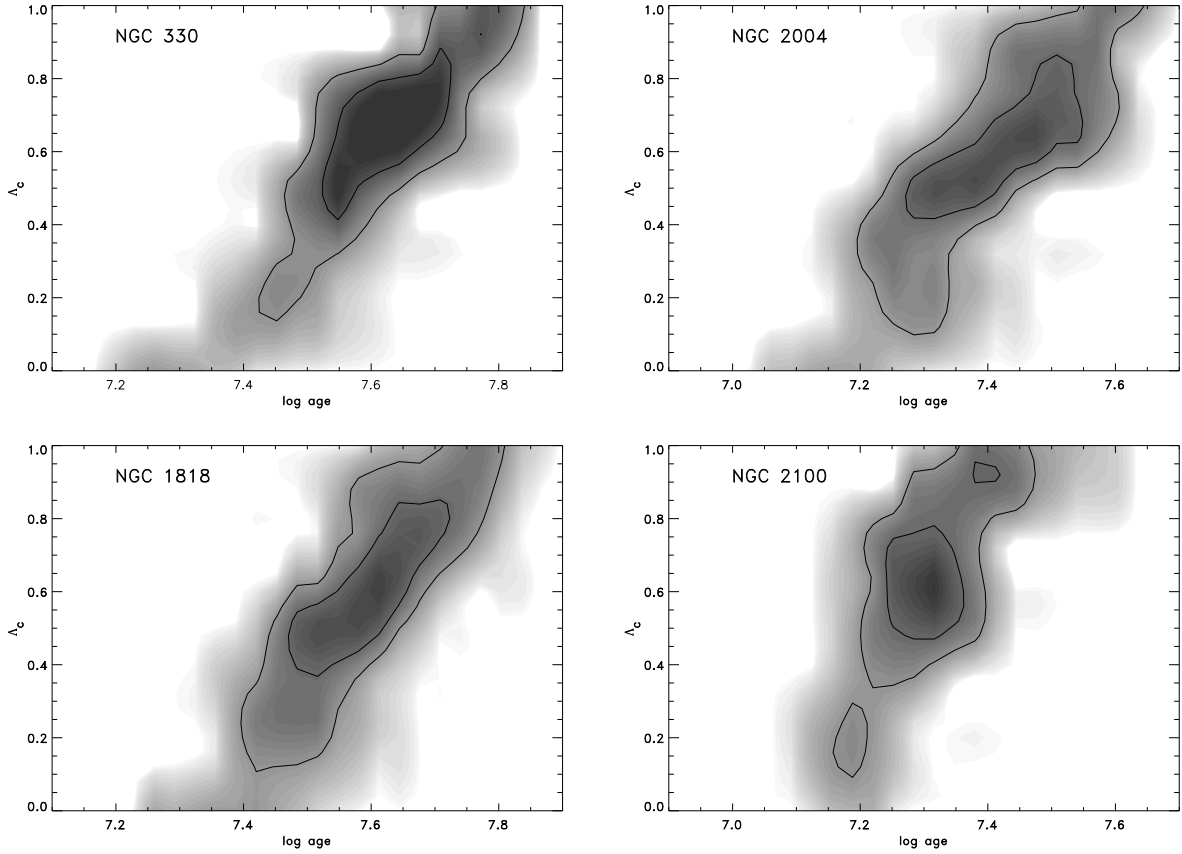}
\caption{Greyscale plot of the joint-probability density from the combined
MSILF and $N_{evol}$ constraints for NGC 330 (top left), NGC 1818 (bottom
left), NGC 2004 (top right) and NGC 2100 (bottom right) with a binary fraction
of 30\%. Contours of 1 and 2$\sigma$ are shown.}
\label{total330bin}
\end{figure*}

\placefigure{total330bin}

Figure~\ref{binnobin} shows a comparison between the MSILF with and without
binaries (30\% as discussed above) for a log age of 7.5 and a mass function
slope of $\alpha$=-2.35. The best-fitting age and $\Lambda_{c}$
for each cluster is given in Table \ref{tab:agebin}. Combination of the
results for the four clusters gives an optimal value of
$\Lambda_{c}=0.62\pm0.22$. 

The inclusion of a significant fraction of binaries alters the MSILF by
flattening the slope of the MSILF - essentially making the model MSILF mimic a
population of younger age without binaries. This has the effect of shifting
the best-fitting model to an older age (Figure~\ref{330bin} (left)). At the
same time, 30\% more stars on the MS shifts the locus of optimal fits to the
number of evolved stars to younger ages (Figure~\ref{330bin} (right)). The net
effect is a slight shift towards higher degrees of overshoot
(Figure~\ref{total330bin} (top left)). Modification of the IMF, on the other
hand, reduces the slope of the MSILF and the number of evolved stars. This
reduces the necessity for CCO but cannot pausibly remove it. From preliminary
modeling of the case of $\alpha$=1.90 (compared with the $\alpha$=2.35
above) reduces the best-fitting $\Lambda_{c}$
but the result remains within the uncertainties detailed above.

%In contrast to our findings, Brocato et al.\ (1989) claim that the presence of
%binaries increases the ratio of the number of MS to post-MS stars and hence
%that the presence of binaries can mimic the presence of CCO and hence reduce
%the necessity for CCO. Brocato et al.\ take the best-fitting model to the LF
%and then superimpose the effects of binaries upon this population. They omit
%to obtain an optimised fit to the LF with the addition of binaries,
%consequently the model population is too young. In our discussion above we
%resample the (age, CCO) parameter space and having done so we find an increase
%in the amount of CCO is required to model the observations in the presence of
%binaries.

\placetable{tab:agebin}

\section{Effective Temperature of the MS}
\label{efftemp}

\begin{figure*}
\plotone{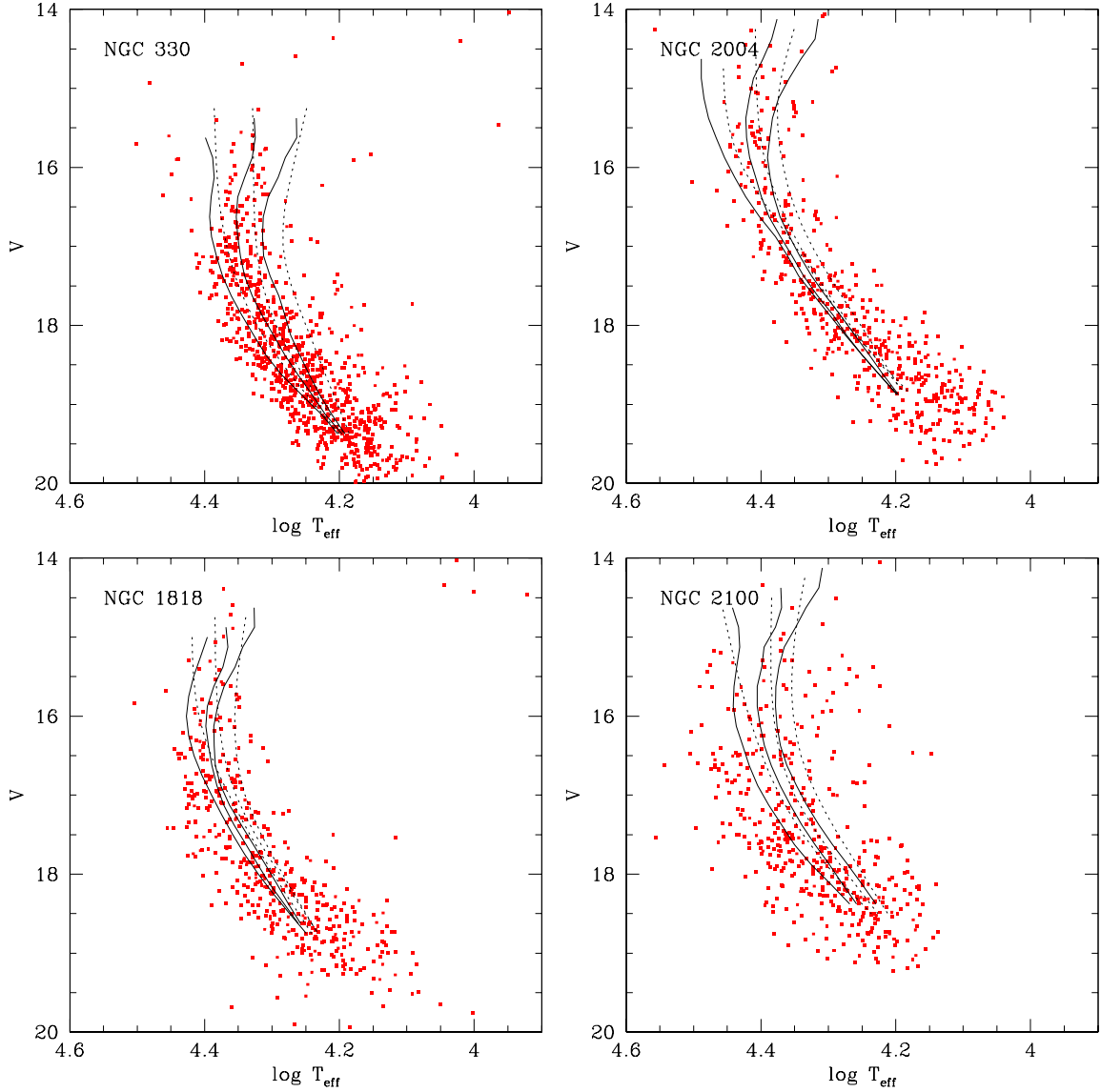}
\caption{$V$, T$_{eff}$ diagrams for the four clusters showing the MS locus
determined from synthetic cluster populations derived for
$\Lambda_{c}$=0,0.5,1 models with the best-fitting ages given in Tables
\ref{tab:agenobin} (solid line - no binaries) and \ref{tab:agebin} (dotted line - with binary fraction). }
\label{teff}
\end{figure*}

As discussed in Paper 1 we have extracted effective temperatures for the MS
population. The temperature information provides a comparison to
the previous determination of $\Lambda_{c}$. Simulations were made for the
best-fitting ages for $\Lambda_{c}$=0, 0.5, 1.0. Simulations were run until
10000 points were generated. Within 0.25 magnitude bins the mean temperature
of each model is shown in Figure \ref{teff}. The addition of a binary fraction
shifts the MS locus towards cooler temperatures and makes the top of the MS
more vertical (dotted line in Figure \ref{teff}). Aside from NGC 2100, which
suffers from differential reddening, the temperature of the upper MS most
closely parallels models with non-zero  CCO.

\placefigure{teffnobin}

\placefigure{teffbin}

\section{Summary} 

Our study of four young populous clusters in the Magellanic Clouds has sought
to quantify the extension to the classical Schwarzschild stellar convective
core by convective core overshoot (CCO). We have compared the
H-R diagrams of each cluster with synthetic ones produced from a set of
stellar evolutionary models different only in their degree of CCO.

Using the twin constraints of (1) the main-sequence LF and (2) the number of
evolved stars, we locate the age and CCO range which maximises the agreement
between theory and observation. In each cluster we find that the case of
no-overshoot is excluded to a 2$\sigma$ level. With the assumption of a
Salpeter mass function and a 30\% binary fraction within the model population, an
average of the results for each cluster yields an overshoot parameter 
$\Lambda_{c}$=0.62$\pm$0.22. A negligable binary fraction favours
$\Lambda_{c}$=0.56$\pm$0.19. Both results are formally in agreement with the
amount of CCO included within the widely used stellar evolutionary models of
the Padua and Geneva groups.  

\begin{acknowledgements} 
SCK acknowledges the support of an Australian Postgraduate Award scholarship
and a grant from the DIST {\it{Hubble Space Telescope}} research fund. We
are very grateful to A. Bressan (et al.) for providing us with the $\Lambda_{c}=1.00$ models. We thank our thoughtful anonymous referee for comments.
\end{acknowledgements}

\clearpage

\begin{table}
\begin{center}
\begin{tabular}{lccc} \hline
Cluster Name & $V_{min}$& $\rm{N_{MS}} (V < V_{min})$   & 
$\rm{N_{Evol}}$ \\ \hline
NGC 330 & 19.50 &  569  &  16  \\ 
NGC 1818 & 18.75&  356  &  12 \\ 
NGC 2004 & 19.00&  426  &  10  \\ 
NGC 2100 & 18.50&  363  &  13  \\ \hline
\end{tabular}
\end{center}
\caption{Summary of our sample. $V_{min}$ defines the $V$ magnitude cutoff for
our sample. $\rm{N_{MS}}$ is the total number of MS stars within the
sample. $\rm{N_{Evol}}$ is the number of post-MS stars within the imaged
field.}
\label{x}
\end{table}

\begin{table}
\begin{center}
\begin{tabular}{ccc} 
\hline 
Model\# & Z & $\Lambda_{c}$ \\ \hline
1& 0.004 & 0.00 \\
2& 0.004 & 0.50 \\
3& 0.004 & 1.00 \\
4& 0.008 & 0.00 \\
5& 0.008 & 0.50 \\
6& 0.008 & 1.00 \\ \hline
\end{tabular}
\end{center}
\caption{The evolutionary models used in the present study.}
\label{models}
\end{table}

\begin{table}
\begin{center}
\begin{tabular}{ccc} 
\hline 
Cluster  & best-fit $\Lambda_{c}$ & best-fit age \\ \hline
%NGC 330  & 0.64 \pm 0.25 & 7.56 \\
%NGC 1818 & 0.51 \pm 0.27 & 7.43 \\
%NGC 2004 & 0.57 \pm 0.23 & 7.29 \\
%NGC 2100 & 0.52 \pm 0.19 & 7.25 \\
NGC 330  & 0.64 $\pm$ 0.24 & 7.56 $\pm$ 0.11\\
NGC 1818 & 0.51 $\pm$ 0.26 & 7.43 $\pm$ 0.12\\
NGC 2004 & 0.57 $\pm$ 0.22 & 7.29 $\pm$ 0.08\\
NGC 2100 & 0.52 $\pm$ 0.20 & 7.25 $\pm$ 0.08\\
\hline 
\end{tabular}
\end{center}
\caption{The best-fitting $\Lambda_{c}$ and age derived for the clusters
assuming no binary fraction.}
\label{tab:agenobin}
\end{table}

\begin{table}
\begin{center}
\begin{tabular}{ccc} 
\hline 
Cluster  & best-fit $\Lambda_{c}$ & best-fit age \\ \hline
NGC 330  & 0.67 $\pm$ 0.20 & 7.62 $\pm$ 0.10\\
NGC 1818 & 0.62 $\pm$ 0.24 & 7.61 $\pm$ 0.13\\
NGC 2004 & 0.61 $\pm$ 0.22 & 7.45 $\pm$ 0.14\\
NGC 2100 & 0.60 $\pm$ 0.16 & 7.31 $\pm$ 0.12\\
\hline 
\end{tabular}
\end{center}
\caption{The best-fitting $\Lambda_{c}$ and age derived for the clusters
assuming 30\% binary fraction.}
\label{tab:agebin}
\end{table}

\clearpage 
% Now comes the reference list.  In this document, we used \cite to call
% out citations, so we must use \bibitem in the reference list, which
% means we use the LaTeX thebibliography environment.  Please note that
% \begin{thebibliography} is followed by a null argument.  If you forget
% this, mayhem ensues, and LaTeX will say "Perhaps a missing item?" when
% you run it.  Do not call us, do not send mail when this happens.  Put
% the silly {} after the \begin{thebibliography}.
%
% Each reference has a \bibitem command to define the citation format
% to be placed in the text (in []) and the symbolic tag used for 
% cross referencing (in {}).
%
% See sample1.tex, or the AASTeX guide, for an alternative to the \cite-
% \bibitem command.

% Option 3. Figures and figure captions are included within figure 
% environments within the body of the manuscript.  In our examples the 
% \plotone command is placed in the figure environment along with the
% figure caption.  The \caption command can also include a \label command.
% Each figure and its caption are printed on the same page.
%
% The \caption command in the figure environment works like the one in the
% table environment (it's the same one, actually), except that this one
% produces identification text that reads "Figure N."
%
% If you wish to see this option then you must comment out all of the 
% \figcaption, \plotone, and \end{document} commands above.

\clearpage

\end{document}